\documentclass[12pt,preprint]{aastex}

\newcommand{\etal}{{\it et al.\/}}

\begin{document} 

\title{Invited Talk: IAU Symp. 266. \\ Globular 
Cluster Abundances and What They \\ 
Can Tell Us About Galaxy Formation}

\author{Judith G. Cohen\altaffilmark{1} }

\altaffiltext{1}{Palomar Observatory, Mail Stop 249-17,
California Institute of Technology, Pasadena, Ca., 91125, 
jlc@astro.caltech.edu}

\begin{abstract}
We review the properties of globular clusters which make
them useful for studying the Galactic halo, Galactic chemical evolution,
and the early stages of the formation of the Milky Way.
We review the evidence that GCs have a chemical inventory similar
to those of halo field stars.  We discuss the abundance ratios
for dSph galaxies and show that it is possible to have formed
at least part
the Galactic halo field stellar population by dissolving globular
clusters and/or accreting dSph galaxies but only if this occurred
at an early stage in the formation of the Galaxy.
We review the constraints on halo formation timescales deduced from the low
Mg isotopic ratios in metal-poor halo field dwarfs
which indicate that AGB stars did not have
time to contribute significantly, while M71 contains two populations,
one without
and also one with a substantial AGB contribution.
We review the limited evidence for GCs with a second
population showing additional contributions from SNII,
currently confined to $\omega$~Cen, M54, and M22, all of
which may have been the nuclei or central regions of
accreted galaxies.  We check our own data for additional
such GCs, and find preliminary indications that NGC~2419, a massive
GC far in the outer Galactic halo, may also belong
to this group.

\end{abstract}


.
\section{Introduction}

Globular clusters (GCs) have unique attributes which make them
laboratories for studying stellar evolution.   Such stellar
systems have well determined distances and ages from their CMDs, accompanied by detailed
kinematic information based on large radial velocity surveys.
They show no evidence for the presence of dark matter, having
$M/L$ ratios expected for old stellar systems.  Their chemical inventories
can be studied at high spectral resolution and with large samples
of stars in a given cluster.  Proper motions are available in some
cases, leading to orbit determinations.  Tidal tails are seen in
a few cases, constraining mass models for the halo. Since
Galactic GCs are old, with typical ages of $\sim$12~Gyr, and
are, at least to first order, formed of stars with identical chemical
compositions, they can provide
unique insights into the formation of the Milky Way Galaxy.

Some of the key issues involving GCs that are of great current interest
are to what extent are there multiple populations in GCs and to
what extent are they chemically homogenous.  Also
whether  the Milky Way halo could have been made
at least in part by dissolving GCs
and accreting dSph satellite galaxies, whether the GCs and the halo have
the same metallicity distribution function (MDF), whether
they have the same chemical inventory, and what we can learn
from them about early SN, star formation, and formation of the
Galactic halo.

The work presented here was carried out in part 
with my former postdoctoral fellows Jorge Melendez
and Wenjin Huang.

\section{Globular Cluster Abundance Ratios vs Halo Field Stars and 
dSph Satellites}

It has been clear for several years now that the abundance ratios
[X/Fe] for Galactic halo field stars show trends with [Fe/H].
Broadly speaking
the ratios increase to a super-Solar level as [Fe/H] decreases
for the $\alpha$-elements, while they decline for Cr~I and Mn~I,
with minimal scatter about the mean relations,
implying that the halo is well mixed with many SN contributing
to the chemical inventory of each star.  A major theme
of work over the past few years has been extending these
relations to ever more metal-poor halo field stars and to exploring
the subtle differences that may exist in chemical inventory
between stars in the the inner and outer parts of the Galactic halo, subjects not part
of the current discussion.  However, over the metallicity range
characteristic of the Galactic GCs, it is now well established
that the GCs, ignoring those associated with the dissolving
Sgr dSph galaxy, follow the trends established by the halo field
stars, see eg the review of \cite{gratton04}.  Detailed chemical
evolution models for the Galaxy such as those of \cite{chem_ev} or
\cite{kobayashi06} can reproduce most of these
trends.  One long-standing issue,
resolved in \cite{hes_mdf} using the MDF
for Galactic halo field stars inferred
from the Hamburg/ESO Survey with corrections for selection efficiency
is that it is probably not statistically significant that there
is no GC with [Fe/H] $< -2.4$~dex given the very low fraction
of extremely metal-poor stars in the halo.

\begin{figure}
\epsscale{0.85}
\plotone{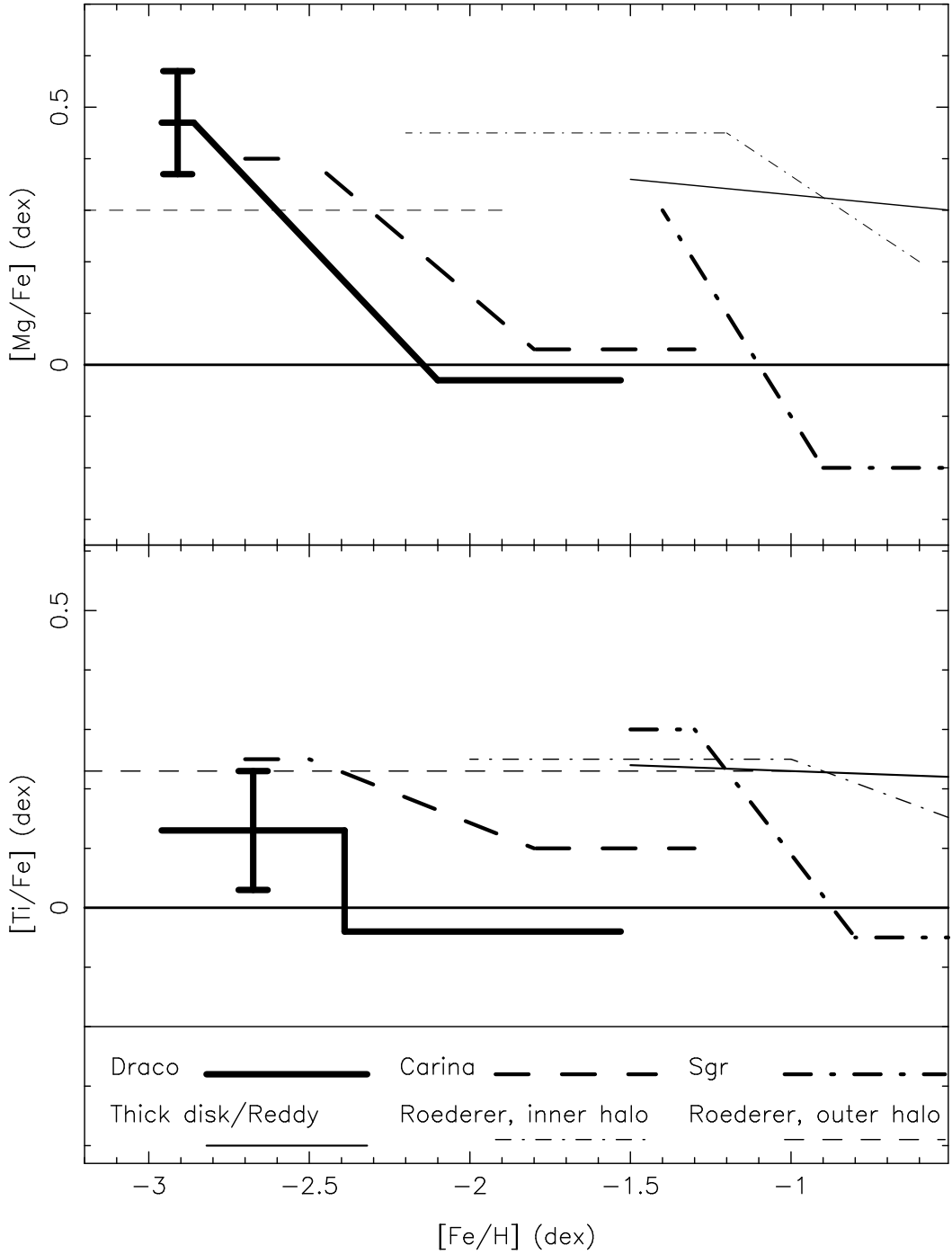}
 \caption{[Mg/Fe] (upper panel) and [Ti/Fe] (lower panel) in three
 dSph satelllites (Draco - Cohen \& Huang 2010, Sgr (Monaco et al 2005;
 Sbordone et al 2007, and Carina - Koch et al 2008)
  as a function of [Fe/H] compared to the trends
 for  Galactic field halo stars from \cite{roederer08} and thick disk
 stars from \cite{reddy06} as well as for Galactic GCs. See Fig.~21 of
 \cite{cohen_draco} for details.}
   \label{fig_draco}
\end{figure}

Another issue is that of the abundance ratios and MDFs for the
satellite dSph galaxies of the Milky Way as compared to the Milky Way
halo.  \cite{helmi06} summarized the position of the very large
DART survey at the VLT, claiming that the dSph galaxies lacked
the expected fraction of extremely metal-poor (EMP) stars, which
are found in the Galactic halo, and that this and other
related evidence from the behavior of abundance ratios
of the $\alpha$-elements precluded the possibility of making
the Galactic halo in whole or part from dissolved dSph galaxies.
This controversial paper has been refuted, and now
even the authors admit that it is flawed.  It is clear
from a large number of recent studies including that of
\cite{cohen_draco} for the classical dSph galaxy
Draco and of \cite{kirby08} for the recently discovered
ultra-faint satellite galaxies that the local dSph galaxies do have
an appropriate (low) fraction of EMP stars, and that at
the lowest Fe-metallicities, [$\alpha$/Fe] in such
low luminosity satellite galaxies is quite close
to the values for similar metallicy stars in the Galactic halo.
At later times, the star formation history of the dSph galaxies
diverged strongly from that of the halo, and strong
differences in chemical inventory and abundance ratios
arose between them.  Thus the bulk of the accretion
must have been early, as is expected from CDM hierarchical galaxy
formation models and simulations.

\section{Elements Known to Vary Within GCs and Why}

The anti-correlation between O and Na is ubiquitous
among GCs.  The range
is very large, sometimes a factor of 10 in abundance of each,
and is seen at all luminosities of the member stars.  There is also
an anti-correlation between Mg
and Al, sometimes with a very large
range in Al (a rather rare element, so burning only a small
amount of Mg into Al will produce a big change in the Al abundance).
GC stars show large star-to-star variations of C and N, almost
always with C and N anti-correlated, see, e.g. \cite{cohen05}.  None of
this is seen among samples of halo field stars.

\begin{figure}
\epsscale{1.0}
\plotone{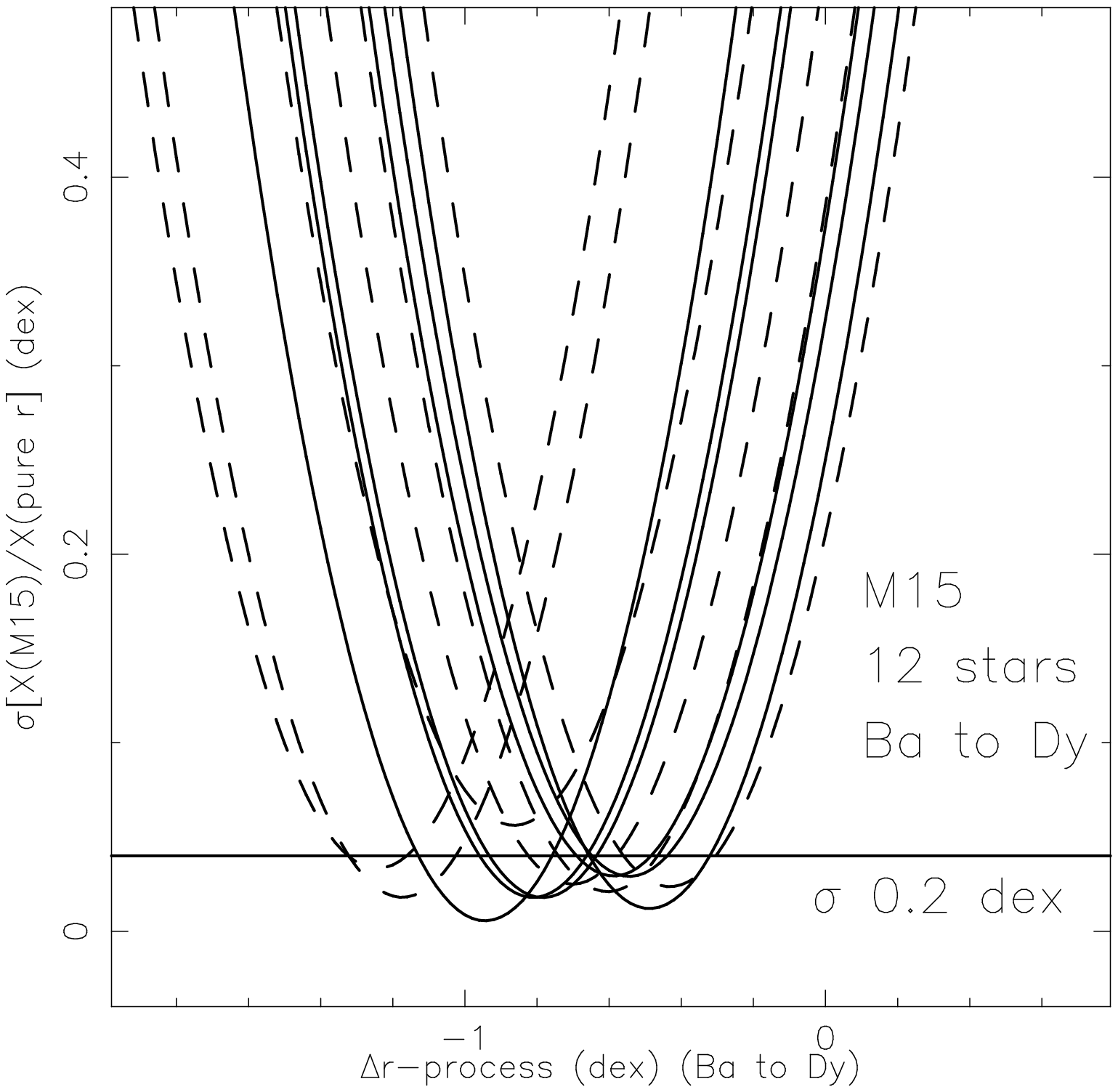}
 \caption{Deviations from a scaled $r$-process are shown for 12 stars in
 the GC M15.  The Y value for each star is the rms of the difference between
 [X/Fe] for the M15 star and that assuming a pure $r$-process distribution
 scaled by [Fe/H] for M15 with a fixed offset ${\Delta}r$
 over the elements between Ba and Dy for which abundances have been derived in the star,
 typically 6 elements.}
   \label{fig_m15_r}
\end{figure}

All of this is tied to HB characteristics, galactic orbits,
almost certainly to He abundances, and is evidence for
at least two generations of star formation in GCs, as is discussed
in a long series of papers by E.~Carretta and collaborators,
the most recent of which is \cite{carretta_nao_8}.

Star-to-star variations of elements between Ca and Ni within GCs  is
discussed in \S\ref{section_sn2}.   Moving on the the heavy
neutron  capture elements beyond
 the Fe-peak,
M15 is the one GC known to show a star-to-star variation among them,
first detected by \cite{sneden97} (see also Otsuki \etal\ 2006).
I (unpublished) 
have verified
 that the variation in abundance among the heavy neutron
 capture elements within M15 is about a factor of 5, and
 the production mechanism for these heavy
 elements is the  $r$-process, rather than the $s$-process.
 Results for a sample of 12 RGB stars in M15 are shown in
 Fig.~\ref{fig_m15_r}.  The low minima in Y
 indicate that the heavy elements in M15 originate in the
 $r$-process.  The range in the value of  ${\Delta}r$
 (the X coordinate of the figure) of the minima for the
 curve for each M15 star
 indicates the range in the fractional $r$-process contribution to the
 chemical inventory of the M15 stars. These are very rare
 elements, $\epsilon$(Eu)/$\epsilon$(Fe) $\sim 10^{-11}$ in M15,
 so small star-to-star variations within a very metal-poor GC should
 not be surprising.

\section{Constraints from Isotopic Abundances}

Isotopic abundance ratios can convey still more detailed information
than atomic abundance ratios as production of a particular isotope
requires  specific nuclear
reactions.  The ratios
among the three stable isotopes of Mg, all of which
can be formed in massive stars 
(\cite[Woosley \& Weaver 1995]{woosley95})
are of particular interst as they provide a rough chronometer.
$^{24}$Mg, the most common
isotope, is formed as a primary isotope from H, while 
$^{25,26}$Mg are formed as secondary isotopes.  The heaviest
Mg isotopes are also produced in intermediate-mass AGB stars
(\cite[Karakas \& Lattanzio 2003]{karakas03}), so the isotopic
ratios $^{25,26}$Mg/$^{24}$Mg increase with the onset of AGB stars.
Therefore, Mg isotopic ratios in halo stars can be used to constrain
the rise of the AGB stars in our Galaxy, and if the AGB contribution
is not present, the minimum timescale for halo formation.

\begin{figure}
\epsscale{1.0}
\plotone{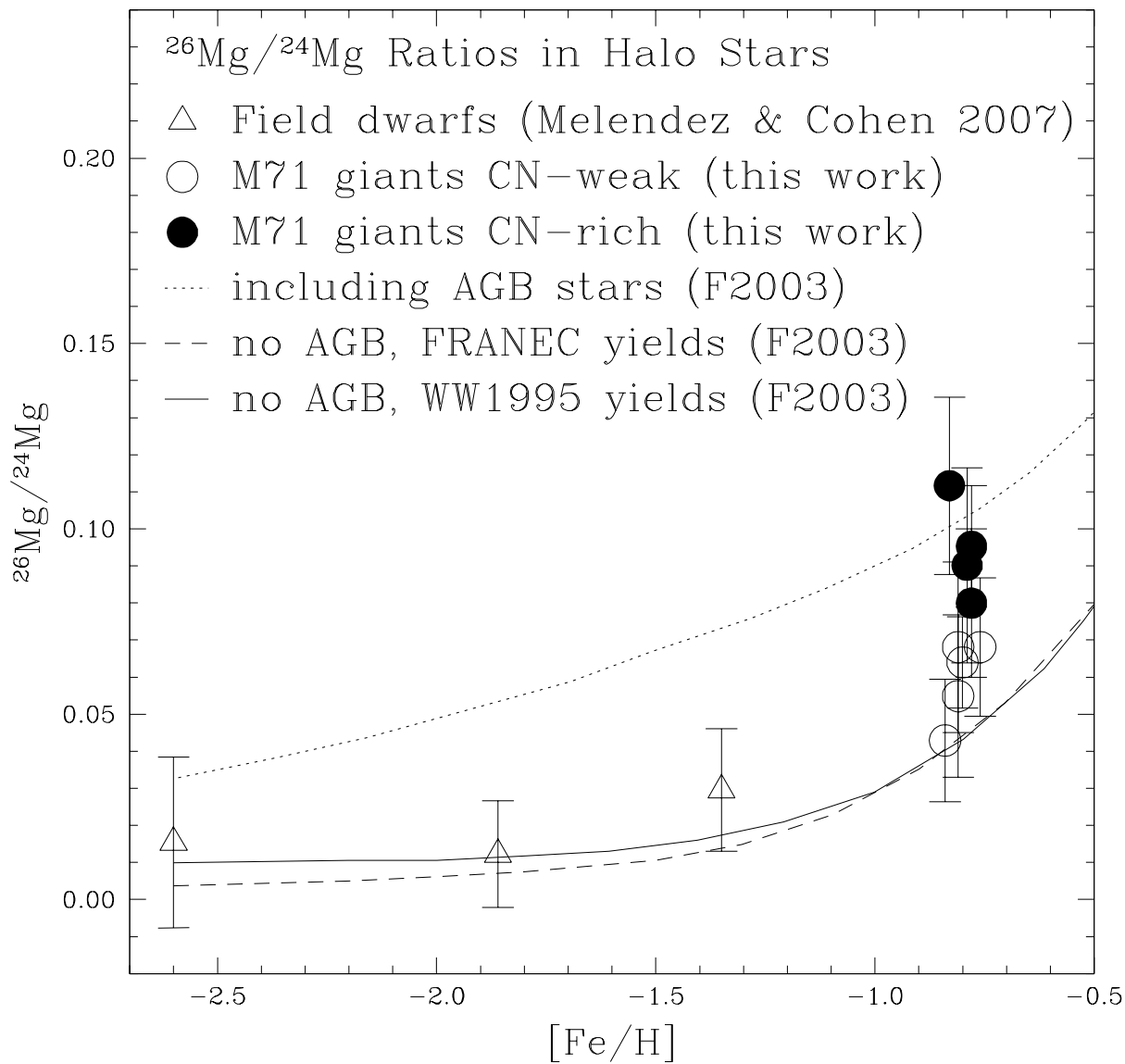} 
 \caption{Our $^{26}$Mg/$^{24}$Mg ratios in both field
dwarfs (triangles; Melendez \& Cohen 2007) and M71 giants (Melendez \& Cohen 2009)
as
 a function of [Fe/H]. 
Chemical evolution models by Fenner et al. (2003) including (dotted line)
and excluding (solid and dashed lines) AGB stars are shown. 
This figure is from \cite{mgh2}.}
   \label{fig_mgh}
\end{figure}

Measurement of the Mg isotopic ratios requires 
detection of weak contributions from the rarer heavier Mg isotopes
in the wings of much stronger
$^{24}$MgH lines, hence exquisite
spectra of very high spectral resolution ($\sim$100,000) and
very high signal-to-noise ratio,
as well as careful attention to details of the line
list used for the spectral synthesis and to the analysis procedure.

This idea is applied in two papers written with my former postdoc,
Jorge Melendez, who did most of the work, \cite{mgh1}
and \cite{mgh2}.   
We see no sign of an AGB contribution
among our sample of
metal-poor
Galactic halo dwarfs.  In the GC M71, at [Fe/H] $-0.7$~dex, 
we see two populations as shown in
 Fig.~\ref{fig_mgh}.  The first has weak CN, normal O, Na, Mg, and Al, and a 
low ratio of $^{26}$Mg/Mg ($\sim$4\%)  consistent with models of galactic chemical
evolution with no contribution from AGB stars.
The second population has
enhanced Na and Al accompanied by lower O and by higher $^{26}$Mg/Mg
($\sim$8\%), consistent with models which do incorporate ejecta
from AGB stars via normal stellar winds.
We therefore infer
that the timescale for formation of the first generation of
stars we see today in this globular cluster must be sufficiently short
to avoid a contribution from AGB stars, i.e. less than $\sim$0.3~Gyr.

\section{Variation of SNII Contributions \label{section_sn2} }

Spectroscopic evidence for multiple populations in Galactic globular
clusters has been known for a long time based on studies
of the O--Na anti-correlation and on spreads in
the CNO abundances among stars within a particular GC  as reviewed
by C.~Charbonnel in this meeting.  These require at least
two generations of globular cluster stars, with the second polluted
by ejecta from either AGB stars or massive rapidly rotating stars.

The existence of multiple generations of stars within GCs has
very recently been detected using photometry as well, as
reviewed in the talk by G.~Piotto in this conference.
The sample of globular clusters with CMDs based on  ACS images taken 
with HST has grown rapidly, most recently through the ACS Survey
of Galactic Globular
Clusters (a HST Treasury project) (\cite[Sarajedini et al. 2007]{sarajedini07}).  
This extremely accurate
photometry has enabled the detection of subtle features in the GC CMDs that
were previously lost in the noise, including 
multiple main sequences, a double
subgiant branch in NGC~1851 (\cite[Milone et al 2008]{ngc1851}), and other such phenomena.

They are usually explained by small differences in age, He content
(which, although difficult to detect directly, must occur in association 
with the O/Na anti-correlation 
and the CNO variations),
or metal content.  When invoking the latter, one must be careful
to distinguish between variations among those elements
such as CNO, which can be
produced by ejecta from intermediate or low mass AGB stars,
and variations in the heavier elements such as Ca or Fe, which
are only produced in SN.  Presumably at such early times, SNII
are the culprit rather than SNIa. Great care is required
to distinguish among the various possibilities.

The issue of a second generation of
SNII contributions is of particular interest, since it is hard
to understand how a low mass GC with low binding energy
could retain energetic SNII ejecta, unless the GC we observe
today were a remnant of an initally much more massive stellar system.
So the question at hand is whether there are any other such cases
in addition to that of $\omega$~Cen, known
for more than 30 years to have a wide
intrinsic range in [Ca/H], [Fe/H] etc.
extending over a range of $\sim$1.3~dex with
multiple peaks (\cite[Norris et al 1996]{norris96}).
The GC M54 also shows this, but it is believed to be part of the central
region of the Sgr dSph galaxy, currently being accreted by the Milky Way.
Very recently M22, under suspicion for many years, was confirmed
to have such by \cite{merino10}, who suggest a range in [Fe/H] of perhaps
$\sim$0.15~dex; see also the talk by J.~Norris
at this meeting.  

Are there more such cases ?  Is this the tip of the iceberg ?
Do other GCs all have more modest variations in [Fe/H] ?
A literature search for  recent detailed
abundance analyses of samples of GC stars, almost always luminous
RGB stars, indicates that $\sigma$[Fe/H] is quite small in
these clusters as compared to the obvious cases described above;
they show dispersions in Fe-abundance of less than 0.05~dex (10\%), and sometimes
as low as 5\%.  
A careful check of a large sample
of GCs, particularly of the most massive of them, is desirable 
as some of the most massive Galactic GCs are not well studied.

\begin{figure}
\epsscale{1.0}
\plotone{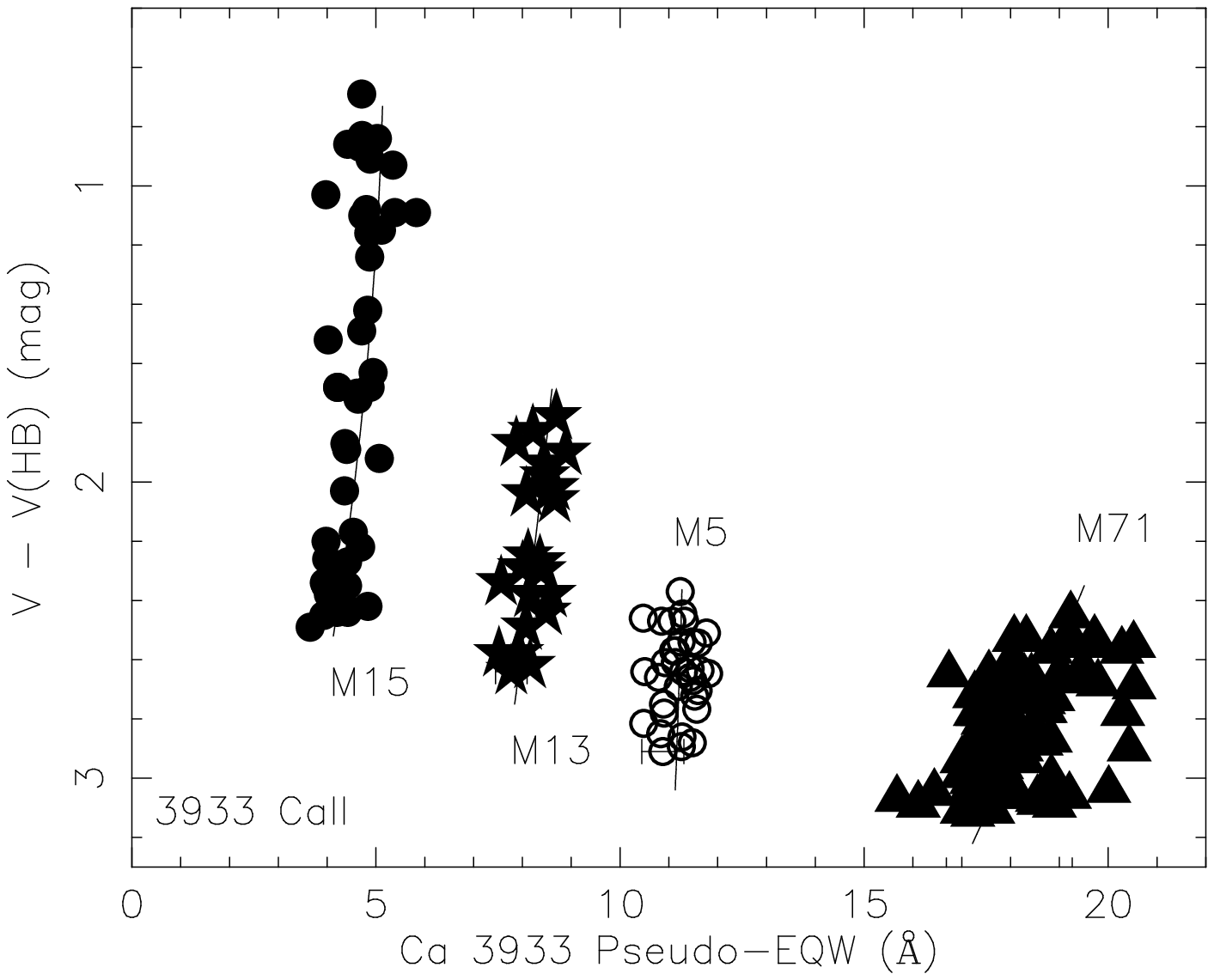}
 \caption{Equivalent width of the 3933~\AA\ line of Ca for samples in four 
 GCs as a function of luminosity below the HB.}
   \label{fig_3933}
\end{figure}

\begin{figure}
\epsscale{1.0}
\plotone{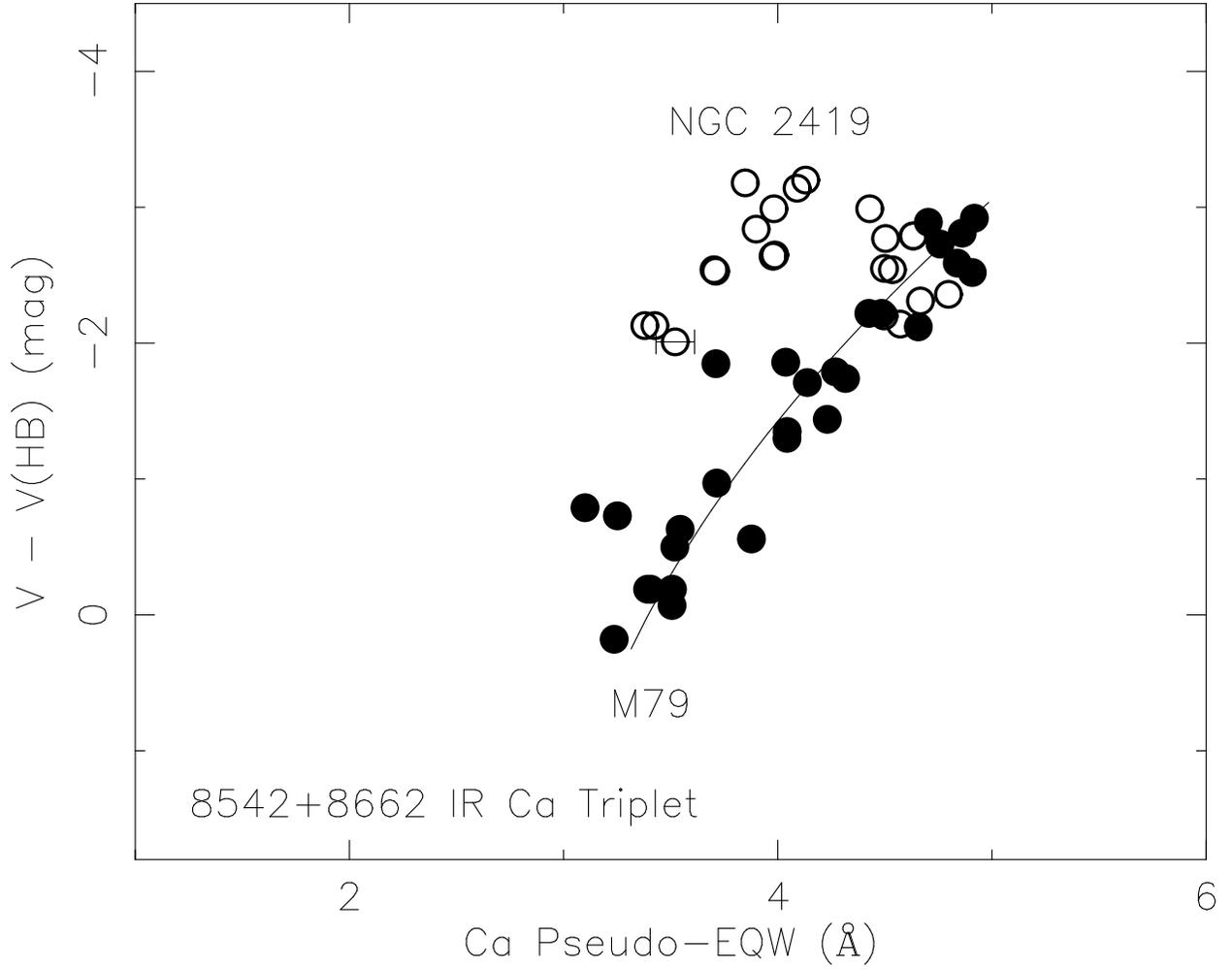}
\caption{Sum of the equivalent width of the 8542 and the 8662~\AA\ lines from
 the Ca triplet for samples of upper RGB stars in M79 and in NGC~2419
 as a function of luminosity.}
   \label{fig_catriplet}
\end{figure}

As a preliminary step towards this goal, I've checked the 
spectra from LRIS/Keck slitmasks I took  from 1998 to 2003
for a study of CH, CN, and NH bands with P.~Stetson
and M.~Briley.  These cover the H and K
lines of Ca~II.  I measured the strength of
absorption in the 3933~\AA\ line.  
Results are shown for four GCs in Fig.~\ref{fig_3933}.
I've also located a substantial number of Deimos/Keck slitmasks
of Galactic GCs which cover the 8500~\AA\ Ca triplet region,
two of which were analyzed prior to the IAU, shown in Fig.~\ref{fig_catriplet}.
These spectra were taken for other purposes and are not optimized
for the present goal, but still yield upper limits to 
$\sigma$[Fe/H] $\sim$0.03~dex in the best cases.

Ignoring M71, which has substantial field star contamination due
to its low galactic latitude,
the most interesting result is that for NGC~2419, a massive GC
far out in the halo at a distance of 84~kpc.
The data in hand appear to show a substantial range of [Ca/H], which
is probably not
due to field star contamination.  Efforts
are now underway both to establish the limiting accuracy that 
can be reached by such techniques
and to verify this interesting result, as, if valid,
NGC~2419 must represent the remnant of a former Milky Way satellite.

\end{document}